\documentclass[epj]{svjour}
\usepackage{graphics}
\begin{document}
\title{Scaling in heavy ion collisions and the low-energy frontier}
\author{Giorgio Torrieri\inst{1} }
%
%
\institute{IFGW Unicamp, Rua Sergio Barque de Holanda 777, Campinas, Brazil}
\date{Received: date / Revised version: date}
%
\abstract{
The common interpretation of elliptic flow $v_2$ in heavy ion collisions is that it is produced by hydrodynamic flow at low transverse momentum  and by parton energy loss at high transverse momentum.   Here, we discuss this interpretation in view of the dependence of $v_2$ with energy, rapidity and system size, and show that it is far from clear how the relevant properties necessary for this interpretation, low viscosity and high opacity, turn on.   A low energy collider such as NICA is essential for this interpretation to be verified, understood and related to the fundamental properties of hadronic matter
\PACS{
      {PACS-key}{discribing text of that key}   \and
      {PACS-key}{discribing text of that key}
     } 
} 
\maketitle
\newcommand{\ave}[1]{\left\langle #1 \right\rangle}
\newcommand{\modulo}[1]{\left| #1 \right|}
\newcommand{\order}[1]{ \mathcal{O} \left( #1 \right) }


\maketitle

Two experimental results of heavy-ion collisions have been subject to many theoretical and phenomenological investigations \cite{sqgp}:  One is the observation of a significant suppression of high transverse momentum ($p_T$) particles, ``jet quenching'', the other one is the observation of an azimuthal dependence of the particle yields on the reaction plane $\phi_{0n}$ at both high and low momenta, the ``harmonic flow''. Harmonic flow
$v_n$ (elliptic flow $v_2$), is defined as the Fourier component of the azimuthal distribution of the produced particles 
\begin{equation}
\label{v2def}
 \frac{dN}{dp_T dy d\phi} =  \frac{dN}{d p_T dy}\left[1+ \sum_{n=1}^\infty 2 v_n(p_T) \cos(\phi-\phi_{0n}) \right]\,.
\end{equation}
The interpretation of the first finding is generally thought to be that the matter produced in heavy-ion collisions is ``opaque'', with a large energy loss per unit length of fast particles \cite{lpm,Baier:1996sk}.
The second finding has been interpreted in terms of the ``perfect fluid'', the hypothesis that matter in heavy-ion collisions has an extremely low viscosity \cite{fluid1,olli}.
Hence, initial anisotropies in configuration-space density of the collision area will be transformed into anisotropies in the collective flow of matter.

$v_n$ is present at all values of momentum, but, if the consensus outlined above is correct, it has different origins at different values of $p_T$.
Specifically, harmonic flow at low-$p_T$ is due to the hydrodynamic evolution of the system, where particles are pushed out differently at different angles due to the azimuthal pressure gradient, while at high-$p_T$ it is due to opacity, since partons emitted in the reaction plane lose less energy than partons emitted perpendicular to the reaction plane due to the shorter distance traveled.  In one case, the driving variable is pressure gradients and particle production happens through something like the Cooper-Frye formula \cite{cf}, and in the other case the driving variable is a path-integrated number of scattering centers, and hadronization, at least at very high $p_T$ should happen via fragmentation functions.  

Both phenomena are thought to be a dynamical response of the primary asymmetry present in heavy-ion collisions.
While the scale delimiting these two regimes is assumed to be the average $p_T$ of the system, $\ave{p_T} \sim \order{0.5-1}$ GeV (with the tomographic regime actually appearing at $\order{3-4}\times \ave{p_T}$), the way these two mechanisms combine at intermediate $p_T$ is not entirely clear.

Phenomenologically distinguishing between different models, even at a given energy, is not so trivial because {\em every} model has quite a few undetermined fit parameters.    Hence, for instance, it is not as yet clear whether jet energy loss proceeds by weakly coupled \cite{lpm,bamps} or strongly coupled \cite{ches1} jet-medium dynamics, and we are far from understanding at what energy, if ever, do these effects disappear.

\begin{figure*}
\begin{center}
  \resizebox{1.\textwidth}{!}{\includegraphics{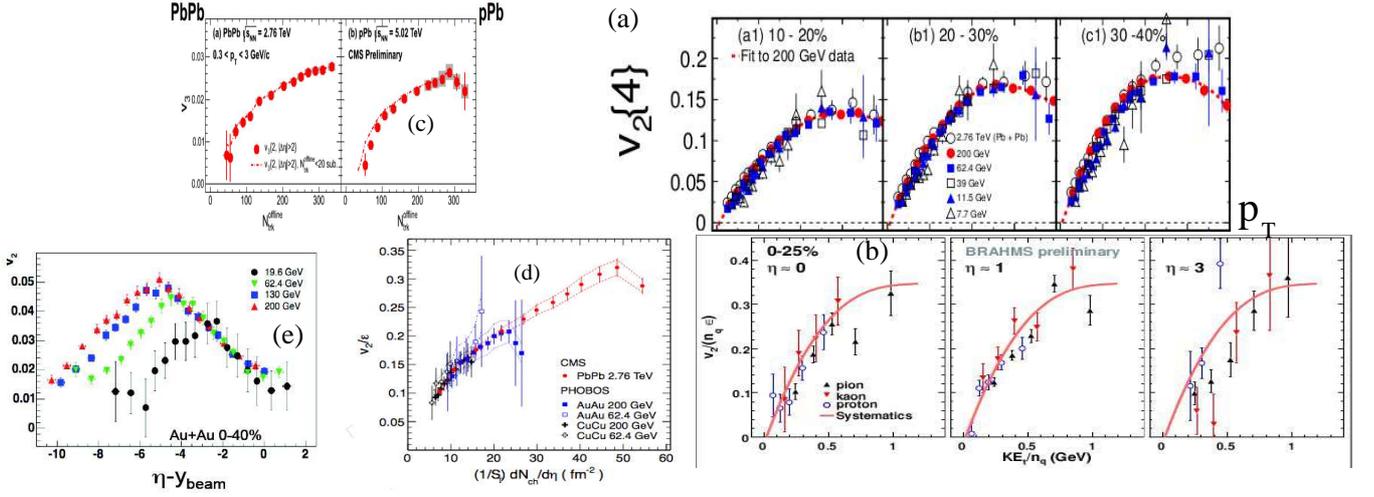}}
  \caption{A compilation of the Experimental puzzles of heavy ion collisions described in this section.   Panels (a,b) show $v_2(p_T)$ independence on energy and rapidity \cite{scanv2paper,brahmsscaling}. Panels
 (d,e,f) show that once $p_T$ was integrated, $v_2$ only depends on the transverse entropy density in the same way as $\ave{p_T}$ \cite{cms,whitephobos}, while (g) shows the same $v_3$ for pA and AA \cite{cmspa}\label{puzzle}}
\end{center}
\end{figure*}

One important experimental finding which can be used to clarify these questions is the discovery of a {\em scaling} in elliptic flow across different energies, system sizes and centralities, when the data is plotted against transverse momentum $p_T$ rapidity $y$,pseudo-rapidity $\eta$ and transverse multiplicity density $(1/S_T)(dN/dy)$ (where $S_T \sim \pi \epsilon_2 R^2$ is the transverse area and $R \sim A^{1/3}$ the size of the system), and eccentricity $\epsilon_n$.   Some experimental observations which help us define this scaling are  (Fig. \ref{puzzle})
\begin{itemize}
\item The dependence of $v_2/\epsilon_2$ on only the transverse multiplicity density $(1/S_T)(dN/dy)$ across all available energies, system sizes and centralities, down to pA and dA collisions and lowest RHIC energies \cite{cms},\cite{phenixda}
\item The ``limiting fragmentation'' of $v_2$ in rapidity \cite{whitephobos}
\item The approximate independence of $v_{2,3}(p_T)/\epsilon_{2,3}$ on energy \cite{scanv2paper}, system size \cite{lacey} and rapidity \cite{brahmsscaling}.
Again, this seems to hold for the lowest RHIC energies as well as, for $v_3$, very small (pA) system sizes \cite{cmspa}
\end{itemize}
One way to parametrize \cite{mescaling0,mescaling1,mescaling2} all this experimental data, at all energies and rapidities, is 
\begin{equation}
\label{scaling1}
v_n(p_T) = \epsilon_n(b,A)F(p_T) 
\end{equation}
\begin{equation}
\label{scaling2}
\ave{v_n} =  \epsilon_n(b,A) \int dp_T F(p_T) \frac{dN}{dp_T}\left( p_T,\ave{p_T}_{y,A,b,\sqrt{s}} \right)  
\end{equation}
\[\  \sim \epsilon_n F\left(\ave{p_T}\right) \phantom{AA} \ave{p_T} \propto \frac{1}{S_T} \frac{dN}{dy} \]
Here, $\epsilon_n$ is the $n$th fourier component of the eccentricity (dependent, in a Glauber parametrization, somewhat weakly on energy, strongly on system size and centrality), and
the distribution in transverse momentum is approximately an exponential with a power-law "tsallis" tail in the high $p_T$ regime,  characterized by one parameter (the average momentum $\ave{p_T}$ or equivalently, the slope $T$), which in turn seem to depend, across rapidity $y$, center of mass energy $\sqrt{s}$ and centrality on just the initial density $\sim \frac{1}{S_T}\frac{dN}{dy}$.
$F(p_T)$ seems to be a universal function, independent of both energy and eccentricity.

These are purely experimental statements, with no theoretical overlay, restating the results \cite{cms,whitephobos,scanv2paper,lacey,brahmsscaling} in mathematical form.  As such, they are ``as good as the error bar'', and a thorough scan in energy, system size and rapidity might in the coming years discover violations.

Taking all this as an established fact, however, is a strong constraint, since observables resulting from a non-linear evolution generally do not factorize: $v_2$ is simply $ v_2(x_i)$ (where $x_i=\left\{ \sqrt{s},b,A,y,p_T \right\}$) and any element of the Hessian matrix is non-negligible \cite{hatta} $$ H_{ij} =  \frac{\partial^2 v_2 }{\partial \ln x_i \partial \ln x_j}\sim \order{1}$$  Eq. \ref{scaling1} implies however that $H_{ij}$ is surprisingly sparse.
This issue is of fundamental importance because the scaling of $v_2$ with energies, system sizes and rapidities is the only all-experimental way to connect the results of heavy ion experiment to variations in the
thermodynamic properties of the system (EoS, viscosity, opacity etc) which are thought to accompany the onset of deconfinement.  In the next two sections we explore what the scalings tell us about the "standard model" of heavy ion collisions, and the role of a future accelerator such as NICA might have in clarifying the situation.
\section{\label{hydro}Scaling in hydrodynamics}
It has long been pointed out, both by heuristic arguments \cite{mescaling1,mescaling2} and explicit simulations \cite{heinzscaling} that the patterns above pose a problem for the hydrodynamic interpretation of $v_2$.
Close to the hydrodynamic limit, one expects that $v_n$ is
\begin{itemize}
\item Approximately $\propto \epsilon_n$ since $v_n (\epsilon_n=0 )=0$ and $\epsilon_n$ is small and dimensionless
\item Approximately $\propto c_s(T)$ where $c_s$ is the speed of sound since $v_2 (c_s=0 )=0$.  
\item $v_2$ is maximum for ideal hydro.   Since the Knudsen number $Kn$, quantifying the ratio of the mean free path to the system size, is small and dimensionless, $v_2 \sim v_2^{ideal} (1-Kn)$.    In turn, the Knudsen number is related to the viscosity over entropy density $\eta/s$ as well as the system size $R$, $Kn \sim \eta/(sTR)$
\item Finally, $v_2^{ideal}$ is a highly non-linear function of the proper time at freezeout $\tau_f$, $v_2^{ideal} \sim v_2(\tau_f/\tau_0 \rightarrow \infty) \times f(\tau_f/\tau_0)$, which can be numerically shown to be monotonically saturating, $\sim  \tanh(\tau_f/\tau_0)$ in a Cooper-Frye \cite{cf,heinzcf} freezeout. $\tau_f$ is in turn related to the freezeout temperature and energy density $T_f,e_f$ while $\tau_0$ is the initial thermalization time, by causality only dependent on local parameters $\sqrt{s},N_{part}^{1/3}$.
\item For small chemical potential and isothermal freeze-out, $\tau_f/\tau_0 \sim (e_{0}/e_f)^{4 \alpha}$, with $\left. \frac{1}{3} \right|_{bjorken} < \alpha < \left. 1 \right|_{hubble}$ depending on how ``three dimensional'' is the flow (in the Bjorken limit expansion is purely longitudinal, in the Hubble limit it is isotropic in all dimensions).  This relation becomes more complicated, but qualitatively similar, for systems at high chemical potential.
\end{itemize}
In summary, elliptic flow in the hydrodynamic limit should scale as
\begin{equation}
\frac{v_2}{\epsilon} \sim c_s  f \left( \frac{1}{T_f^3 \tau_0 R^2}\frac{dN}{dy}  \right) \left( 1 - \order{1} \frac{\eta}{s}  \frac{1}{ T R} \right) 
\label{scaldn} 
\end{equation}
The analytical solution in \cite{hatta} gave somewhat different scaling, most likely due to the assumption that the freeze-out time is parametrically smaller than the system size.   This assumption was necessary for the consistency of the analytical solution but it is not supported by data, and dimensional analysis arguments \cite{hatta} show that the scaling of Eq. \ref{scaldn} is recovered once a later freezeout is assumed.  Note that in these reasoning we {\em optimize} scaling by neglecting temperature dependent $\eta/s$, which is physically reasonable and will surely complicate things \cite{etasT}.

It is clear that only $\order{Kn}$ terms mix intensive quantities such as the energy density $e$ with extensive ones such as the size $R$.    $\order{Kn^0}$ ``ideal'' terms, except for the initial time $\tau_0$, depend purely on intensive quantities, giving rise to scaling between systems of different sizes.   As we will see, this is {\em not} true for high $p_T$.
For low $p_T$, no scaling violation is seen in experimental data, giving a bound for the Knudsen number compatible with $\eta/s=0$ \cite{nantes1}, albeit with large error bars (which still need to be computed), at all energies.

Moreover, the lack of scaling of $\tau_0$ is troubling since, by causality, it can locally {\em only} depend on energy $\sqrt{s}$, thickness function $A^{1/3}$ and the local intensive parameters.
Just by dimensional analysis, it is difficult to see how   It can be constant w.r.t. energy, since $\tau_0 \sqrt{s} \sim \order{1}$.  Landau hydrodynamics would imply $\tau_0 \sim 1/\sqrt{s}$, while a CGC type initial condition would most likely give a logarithmic dependence since $\tau_0 \sim Q_s^{-1}$ where $Q_s$ is the saturation scale.   Either, however, would lead to unobserved systematic scaling violations.   The only way to avoid these is to assume $\tau_0$ is of the order of the local mean free path at equilibrium, and hence gets reabsorbed as a function of the entropy density, $\left( \sim (1/S_T)(dN/dy) \right)^{-1/3}$ for an ideal Equation of state.

Additionally, the Cooper-Frye formula \cite{cf,heinzcf} leads to a {\em non-universal} $F(p_T)$.    To show this, it is sufficient to expand it azimuthally in eccentricity \cite{cf,heinzcf}
 \begin{equation}
v_2(p_T) \simeq \int d\phi \cos^2 (2\phi) 
\end{equation}
$$\left[  e^{ - \frac{\gamma\left(E-p_T  u_T \right)}{T}} \left(1 - p_T \Delta \frac{dt}{dr}  +  \frac{\gamma \delta u_T (\phi)  p_T }{T}  + \order{\epsilon^2,Kn} \right) \right]$$
As long as $\delta v_T,  \Delta \frac{dt}{dr}  \sim \epsilon s^0$, $v_2(p_T)$ is independent of $\sqrt{s}$.   This will be true in the limit where the hydrodynamic phase is ``long'', $\tau_f/\tau_0 \gg 1$, but will {\em not} be the case \cite{heinzscaling,olli} if the duration of the hydrodynamic phase $ \leq \epsilon R$, as is the case at lower energies.   The introduction of an ``iso-knudsen freezeout'' rather than an isothermal one, a physically reasonable scenario explored in \cite{duslingteaney}, should further break this scaling.
The reasons for this behavior go all the way to the qualitative description of how $v_2$ behaves in hydrodynamics: $v_2(p_T)$ and $p_T$ integrated $\ave{v_2}$ scale differently, because in hydro the Fourier components of the transverse flow
$v_T(r,\phi)$ depend on lifetime $\tau_f$ differently:
\begin{description}
\item[$v_2(p_T)$] depends only on the 2nd Fourier component of $v_T(r,\phi) \sim \tanh(\tau_f/\tau_0)$
\item[$\ave{v_2}$] depends on both the 0th ($\ave{p_T}\sim  \left( \tau_f/\tau_0 \right)^\omega$) and 2nd component.   
\end{description}
Given these, making $v_2(p_T)$ independent of energy but varying $\ave{p_T}$ strongly at all energies is unnatural in hydrodynamic models, something that has been confirmed with explicit simulations \cite{heinzpt,sorensen}.  Detailed simulations including chemical potential \cite{petersen,denicol}, however, are needed to determine when does $\tau_f$ become ``short'', and more experimental data might yield a breakdown of $v_2(p_T)$ scaling at lower energies.

A further consideration is in order regarding the breakdown in scaling in {\em particle species} \cite{scanv2paper}.
$v_2^i(p_T)$ ($i=\pi,K,p,\Lambda,...$) does {\em not} scale the same way by particle species as it does for all particles:   different particle species $v_2^i(p_T)$ are different at different energies, but the differences cancel out when the total $v_2 \simeq \sum_i v_2^i(p_T) n_i(p_T)/(\sum_i n_i(p_T))$ is considered ($n_i$ is the particle species abundance).
In a Cooper-Frye freezeout \cite{cf} there is no reason for such a cancellation between flow and hadrochemistry to happen.   Coalescence models, while they will also break coalescence {\em scaling} at lower energies \cite{dunlop,megreco}, also do not predict such behavior.   

\section{Scaling in the tomographic regime\label{jets}}
Unlike in the hydrodynamics flow, where $p_T$-correlations are generated by density {\em gradients}, in tomography such differences are generated by path length variations.   Hence, the role of ``size'', $\ave{R}$, which typically depends on system size as $\sim A^{1/3}$ and is weakly dependent on energy, is very different in the tomographic regime w.r.t. the hydro regime. As we saw in the previous section, ``extensive'' factors $\ave{R T}$ (size$\times$ temperature) in the hydrodynamic regime are suppressed by $\order{Kn}$, and hence vanish in the ideal hydrodynamic limit.   In the tomographic regime, $v_2 \sim \epsilon f(\ave{R T} )$ and the $\ave{R T}$ dependence is not suppressed by any small parameter.   The exact form of the function depends on the details of the model, but, as we shall see, it generally does not drop out when $\ave{R}$ and $\ave{T}$ are varied separately by scanning in centrality, system size and $\sqrt{s}$.     This is true in ``standard'' jet energy loss calculations, both weakly \cite{lpm,Baier:1996sk} and strongly coupled \cite{ches1}, where opacity is a smooth function of the entropy density, but the scaling violation should be enhanced in models such as \cite{betz3}, well-founded in QCD, where opacity is non-monotonic w.r.t. temperature.

The ABC model is a simple parametrization which describes the energy loss of a ``fast'' particle ($p_T/T \ll 1$) in traveling ``large'' medium ($1/(T \tau) \ll 1$, $\tau$ is the propagation time).   If the parton is light and on-shell, energy loss models should give
\begin{equation}
-\frac{dE}{d\tau} = f(T,p_T,\tau) \simeq \kappa p^{a} T^{b} \tau^c + \order{\frac{T}{p_T},\frac{1}{T\tau}}
\label{abceq}
\end{equation} 
$a,b,c$ are a nice  phenomenological way of keeping track of \underline{every} jet energy loss model in certain limits.   
In a collisional dominated parton cascade $a=1,c=0$ \cite{bamps}, in a radiative dilute plasma (``Bethe-Heitler regime'') $a=1/3, c=0$, in a dense plasma (LPM regime) $a=1,c=1$) while in a ``falling string'' AdS/CFT scenario \cite{ches1} $a=1/3,c>2$.   

We can now expand in ``empirical'' parameters $\left( \frac{\Delta E}{E}\right)^{\pm 1},\epsilon,$ and get
\begin{equation}
 v_2\left( p_T^{high} \right)  \sim \epsilon^{\alpha_{\epsilon}}  L^{\alpha_R(a,b,c) } p_T^{\alpha_{p_T}(a,b,c) } \left( \frac{dN}{dy} \right)^{\alpha_{dn/dy}}.
\end{equation}
with non-trivial exponents $\alpha_i$ necessitating a systematic pQCD treatment beyond the scope of this white paper.
The size parameter $L \sim \min\left( \ave{R},\tau_f-\tau_0\right)$, sampling the lifetime for some systems and the transverse size $\ave{R}$ for others.
We perform a numerical investigation with a background given by a longitudinally expanding ellipsoid, fitted to global multiplicity and system size
\begin{equation}
T(r_T,\phi,\tau,\tau_0) = T_0 \left(\frac{\tau_0}{\tau}  \right)^{1/3} \Theta\left( r - R\left( 1+ \epsilon \cos\left(2 \phi \right) \right) \right) 
\end{equation}
Where $R,\epsilon$ are scanned scanned across radii of Cu,Au,Pb, $\tau_0$ is chosen according to the assumptions described in section \ref{hydro} $\tau_0= 1fm,\sqrt{s}^{-1}, T_{0}^{-1}$, and $T_0$ is adjusted to reproduce multiplicity and all energies.    We then obtain $v_2(p_T)/\ave{v_2}$, the latter given by the experimental parametrization $\ave{v_2}/\epsilon = 4(10^{-3}) (1/S_T)(dN/dy)$.
Since the reaction plane was determined at low $p_T$, one can now eliminate the theoretical input $\epsilon,S_T$ from our observables by concentrating on $v_2(p_T)/\ave{v_2}$.   The denominator is the momentum-integrated $v_2$, which should include all eccentricity information.    If $v_2$ depends purely on gradients, of course, this ratio should be independent of both energies and system sizes.  The size dependence, however, should lead to a break in the scaling, in all models.

\begin{figure}
  \resizebox{0.4\textwidth}{!}{\includegraphics{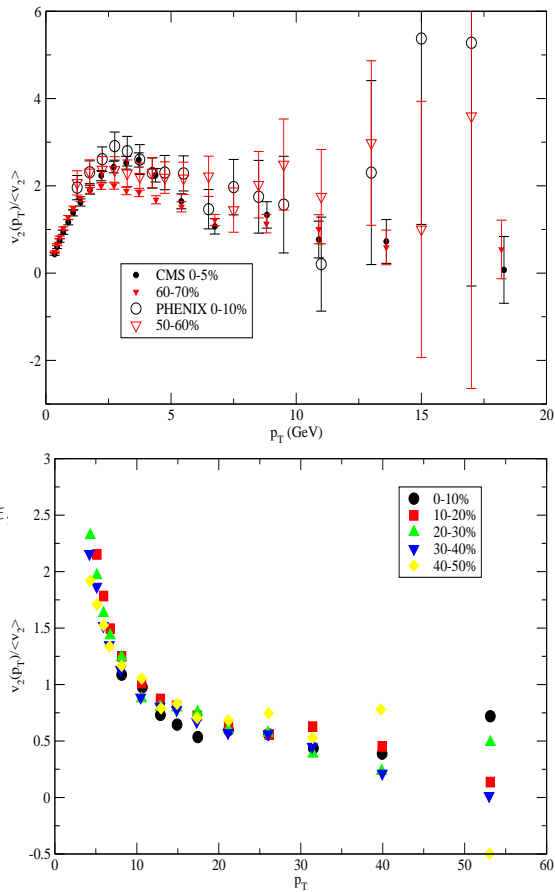}}
  \caption{\label{cmsexp} Experimental $v_2(p_T)/\ave{v_2}$ for LHC and RHIC energies.  The top panel shows data from \cite{cms,lacey} at comparatively low $p_T$, the bottom panel from \cite{cmshighpt}.}
\end{figure}

Currently, experimental data do not allow to make definite conclusions but, as shown in Fig. \ref{cmsexp} \cite{cms,cmshighpt,lacey}, modulo rather big error bars, the only scaling violation is seen at intermediate $p_T$.    Scaling holds across different centralities up to well above $p_T \simeq 20$ GeV, and seems to break up only at $p_T \simeq 40$ GeV at the LHC.  While a systematic shift of the center is seen comparing RHIC and LHC, the error bars are way too big to attach any meaning to this conclusion.  This shift is {\em not } seen up to $p_T \simeq 3.5$, where, as  noted in \cite{lacey}, this scaling holds across RHIC Cu+Cu and Au+Au.

\section{The necessity of a low energy collider}
What are we to make of all these scalings?  Theoretically we believe they represent a challenge to theory which is yet to be fully addressed.
The simplest hypothesis, in the framework of the "standard model of heavy ion collisions", is that viscosity, EoS and opacity are similar across all energies, system sizes and rapidities.   Since in hydrodynamics the scaling violation effects are subleading in Knudsen number, and in tomography these effects depend on variations in temperature that are logarithmic, {\em provided the mechanisms for generation of hydrodynamic and tomographic $v_2$ remain unchanged}, it is possible that at {\em all} RHIC and LHC energies $v_2$ is simply "saturated" at a value regulated by Eq. \ref{scaling1} and \ref{scaling2}.

If this is the case, what can we expect at {\em low} energies?   The compilation in \cite{petersen} (Fig 28,29) makes it clear that Eqs \ref{scaling1} and \ref{scaling2} {\em cannot} possibly hold in a regime where $\sqrt{s} \sim 1$ GeV, because there $v_2$ is strongly non-monotonic.   How, in this non-monotonic regime $v_2(p_T)$ behaves, using different definitions of $v_2$ (reaction plane, cumulants, L-Y zeroes) has never been investigated and, as section \ref{hydro} makes clear, the non-monotonicity observed there should be directly related to non-monotonicity of parameters like the speed of sound and $\eta/s$.
To investigate this experimentally, one needs to perform high-statistics low energy measurements of $v_2(p_T,y)$ at both different energies and system sizes ($pA$ and different kinds of $AA$).    If hydrodynamics is indeed the origin of $v_2$, a scan in $p_T$ and system size at low energy will disentangle variations of viscosity, speed of sound and lifetime.

 Regarding opacity, it should be noted that current measurements of jet energy loss at low energy, such as \cite{starcp} Fig 2 tell us very little about actual opacity since at lower energies, just from kinematics, the Cronin effect starts playing an overwhelming role (sequential scattering is needed to produce a particle of $p_T \geq \sqrt{s}$ in nuclear collisions).   
$v_2$ in the tomographic regime of course continues to be a good observable.   In particular, it is independent of effects like Cronin effect and jet reconstruction, which become problematic at low energies:   At lower energies $R_{AA}$ diverges, due to kinematics, at {\em any} medium opacity.   Since kinematic effects, {\em by themselves}, do not depend between the hadron-reaction plane angle, however, $v_2$ does need opacity to be generated.    This makes comparing $v_2 (p_T>p_T^{tomo})$ at {\em different energies and system sizes}, and looking for scaling violations, an optimal probe of changes in opacity with temperature.
Hence,  The only way to measure opacity at low $\sqrt{s}$ is to measure $v_n$ at high $p_T$ and low energy, a very high statistics measurement which
nevertheless depends on opacity only.    To do so one must perform the sort of analysis described in section \ref{jets}, necessitating data from the same detector in both low and high $p_T$.    

One thing that is of huge experimental help is that when we go lower in energy, the definition of "tomographic" w.r.t hydrodynamic also changes.
Generally, the transition between ``hydrodynamics'' and tomography is a smooth superseding rather than a ``turning on/off''.   Indeed, hydrodynamics and tomography can be defined in terms of the Knudsen number in momentum space:
 assuming the scattering cross section $\sigma$ depends on the exchanged momentum $Q$ as $\sigma \sim 1/Q^2$, and assuming momentum is much higher than temperature, the Knudsen number becomes
$\frac{l_{mfp}}{R} \sim \frac{p_T^2}{s R}$
Hence, the tomographic regime starts dominating when $Kn \sim 1$. For different energies and transverse densities the critical $p_T$ can be easily shown to be
\begin{equation}
 \frac{p_{T1}^{tomo}\left( \sqrt{s_1},b_1,A_1 \right)}{p_{T2}^{tomo}\left( \sqrt{s_2},b_2,A_2 \right)} \sim \left( \frac{S_2}{S_1} \frac{dN_1/dy}{dN_2/dy} \right)^{\omega}
\end{equation}
where $\omega=1/2$ for collisional-dominated equilibration but could increase to $3/2$ \cite{dusling} if radiative processes become important in the hydrodynamic regime.
The advantage of characterizing as ``hard'' hadrons with  $p_T \geq p_T^{tomo}$ is that this definition is independent of details such as the dynamics of production and fragmentation of fast hadrons (we do not have to call them ``jets'', which is problematic at low energies).   Experimentally, the fact that $p_T^{tomo}$ decreases with decreasing $N_{part},\sqrt{s}$ is advantageous.   Measuring the energy and size dependence of $v_2(p_T^{tomo})/v_2(p_T^{hydro})$ should highlight any change in either opacity or mechanism of particle energy loss.

These measurements are more appropriate for collider than for fixed target experiments.  The fact that $v_n$ hinges on meaurements of multi-particle azimuthal correlations and higher cumulants (2,4,6,8 particle cumulants and L-Y zeroes \cite{v2methods}) makes detectors with high acceptance symmetry essential.  Collider experiments are the only experiments that can deliver such symmetry together with energy and system size flexibility, where $\sqrt{s}$ can be tuned at will to look for scaling violations.

In conclusion, we highlighted the remarkably simple scaling of certain crucial observables in heavy ion collisions, and pointed out that this very simple scaling shows some tension with what is considered to be the standard model of heavy ion collisions.   However, we also pointed out that this scaling is almost certain to break at lower energies, and advocated extensive energy scan measurements at a future low energy heavy ion collider such as NICA as a crucial probe to detect any non-monotonicies in how parameters responsible for collective dynamics, such as viscosity and opacity, evolve with initial temperature.   Without such a detection, we are afraid the onset of deconfinement will never be experimentally ascertained.

GT acknowledges support from FAPESP proc. 2014/13120-7 and CNPQ bolsa de
produtividade 301996/2014-8.



\bibliographystyle{aipproc}   





\end{document}